\documentclass[11pt, oneside]{article}   	
\usepackage{amsmath,amssymb,amsthm,geometry,hyperref}
\usepackage{graphicx}						
\usepackage{amssymb}
\usepackage[utf8]{inputenc}
\usepackage[T1]{fontenc}

\newcommand{\be}{\begin{equation}}
\newcommand{\ee}{\end{equation}}
\newcommand{\bea}{\begin{eqnarray}}
\newcommand{\eea}{\end{eqnarray}}
\newcommand{\bean}{\begin{eqnarray*}}
\newcommand{\eean}{\end{eqnarray*}}
\newcommand{\nn}{\nonumber}

\title{Closed-Form Solutions of Zero Dimensional $\phi^4$-Field Theory Using Bessel Functions: A Non-Perturbative Approach}
		
\author{Ranjiva Munasinghe}

\begin{document}

\maketitle	

\begin{abstract}
The integral $\int_{-\infty}^{\infty} e^{- x^2 - g x^4} dx $ is used as an introductory learning tool in the study of Quantum Field Theory and path integrals. Typically it is analysed via perturbation theory. Close form solutions have been quoted but it is not clear how they were derived. So I set about deriving the close form solution on my own and using the same methodology obtain closed form expressions for the even positive integer moments. 
\end{abstract}

\section{Introduction}
Path integrals, Feynman diagrams and perturbation theory are well-established tools in the study of Quantum Field Theory (QFT). A typical introduction to these techniques is via "toy model" 
\be
\label{eq:main_integral1}
Z(g) = \frac{1}{\sqrt{\pi}} \int_{-\infty}^{\infty}  e^{- x^2 - g x^4} dx \;\;\; g \geq 0.
\ee
The function $Z(g)$ is referred to as the partition function of i) {\it the zero dimensional anharmonic oscillator} \cite{HK} or alternatively ii) {\it zero dimensional $\phi^4$-field theory} \cite{JCC, Yuk}.  Analyzing the behaviour of (\ref{eq:main_integral1} is done via deriving perturbative expansions for $Z(g)$ in (\ref{eq:main_integral1}). This is done by expanding the exponential in the integral (\ref{eq:main_integral1}) and interchanging the order of the resulting summation and integration. In the {\it weak coupling} limit $g \rightarrow 0$ one obtains the divergent asymptotic expansion \cite{JCC, HK, Yuk}
\be
\label{eq:series1}
Z(g) \sim \sum_{n=0}^{N} (-1)^n \frac{ \Gamma (2n + 1/2)}{n! \sqrt{\pi}} g^n.
\ee
In the {\it strong coupling} limit $g \rightarrow \infty$ we obtain the convergent expansion  \cite{HK}
\be
\label{eq:series2}
Z(g) \sim g^{-1/4} \sum_{n=0}^{N} (-1)^n \frac{ \Gamma (n/2 + 1/4)}{2n! \sqrt{\pi}} g^{-n/2}.
\ee
The references \cite{HK, Yuk} state (without derivation) the closed-form solution to the integral (\ref{eq:main_integral1}):
\be
\label{eq:main_soln1}
Z(g) = \frac{1}{\sqrt{4 \pi g}} \exp \left[\frac{1}{8g}\right]K_{1/4} \left( \frac{1}{8g} \right).
\ee 
Note the {\it modified Bessel function of the second kind} $K_{\nu}$, also called a MacDonald function \cite{Yuk}, in (\ref{eq:main_soln1}). The function can be expressed for $|\arg z| < \pi/2$, i.e. Re $z > 0$, as \cite[p. 181]{GNW}:
\be
\label{eq:KBess}
K_{\nu}(z) = \int_{0}^{\infty} \cosh (\nu t) e^{- z \cosh t} dt.
\ee
The expression (\ref{eq:main_soln1} is in agreement with the formula in \cite[Ch. 3.323 No. 3 p. 337]{GRI}, which in turn refers to the formula in \cite[Ch. 4.5 No. 34 p. 147]{ERD}. No derivation of the formulae are stated in either \cite{ERD, GRI}. In addition we note the alternative formulation of the closed-form solution for $Z(g)$ in \cite{Yuk}:
\be
\label{eq:main_soln5}
Z(g) = \left( \frac{1}{2g} \right)^{1/4} \exp \left[ \frac{1}{8g} \right] D_{-1/2} \left( \frac{1}{\sqrt{2g}} \right).
\ee
The equation (\ref{eq:main_soln5}) can be verified from (\ref{eq:main_soln1}) using the identity \cite{Wolf}:
\be
D_{-1/2}(z)=	\sqrt{\frac{z}{2 \pi}} \cdot K_{1/4} \left( \frac{z^2}{4} \right).
\ee
As a final remark in this background section, we note that both expansions, (\ref{eq:series1}) and (\ref{eq:series2}), can also be obtained from (\ref{eq:main_soln1}) using the appropriate expansion of $K_{\nu}(z)$ \cite{Yuk}. 

\section{Derivation of Closed Form Expression for Partition Function}
We first prove the formula in \cite[Ch. 3.323 No. 3 p. 337]{GRI}. Using the same convention in \cite{GRI} we consider the integral
\be
\label{eq:main_integral2}
I(\gamma, \beta) = \int_{0}^{\infty}  \exp \left[ - 2\gamma^2 x^2- \beta^2 x^4 \right] dx
\ee
We are interested in the setting $\beta, \gamma \in \mathbb{R}$. We should remark that in the formula \cite[Ch. 3.323 No. 3 p. 337]{GRI} is quoted to hold under the more general conditions $| \arg \beta | < \pi/4$ and $| \arg \gamma | < \pi/2$. We define a new variable transformation
\be
\label{eq:xtrans1}
x = \frac{\sqrt{2}\gamma}{\beta} \sinh \left( \frac{\xi}{4} \right).
\ee
Using (\ref{eq:xtrans1}) (please refer details in Appendix \ref{sec:appA}) and the definition (\ref{eq:KBess}), we re-write ((\ref{eq:main_integral2}) as
\bea
I(\gamma, \beta) &=& 2^{-3/2} \frac{\gamma}{\beta} \exp \left[ \frac{\gamma^4}{2\beta^2}  \right] \int_{0}^{\infty} \cosh \left( \frac{\xi}{4} \right) \exp \left[ - \frac{\gamma^4}{2\beta^2}  \cosh \xi \right] d\xi \nn \\
	&=& 2^{-3/2} \frac{\gamma}{\beta} \exp \left[ \frac{\gamma^4}{2\beta^2} \right] K_{1/4} \left( \frac{\gamma^4}{2\beta^2}  \right) \label{eq:main_soln2}
\eea
Equation ((\ref{eq:main_soln2}) is the formula  \cite[Ch. 3.323 No. 3 p. 337]{GRI}. To verify (\ref{eq:main_soln1}) we substitute $\gamma^2 = 1/2$ and $\beta^2 = g$ in (\ref{eq:main_soln2}) and use:
\be
Z(g) = \frac{2}{\sqrt{\pi}} \cdot I\left(\pm \frac{1}{\sqrt{2}}, \pm \sqrt{g}\right).
\ee

\section{Closed Form Expression for The Moments}
Let us return to partition function (\ref{eq:main_integral1}). We note that the integrand is an even function and should thus only have even (positive integer) moments given by the formula:
\be
\label{eq:main_integral3}
\biggl< x^{2n} \biggr> = \frac{1}{Z(g)} \cdot \frac{2}{\sqrt{\pi}} \int_{0}^{\infty}  x^{2n} \exp \left[- x^2 - g x^4 \right] dx
\ee
To obtain our closed form expression we proceed as before, now using the variable transformation
\bea
\label{eq:xtrans4}
x &=& \frac{1}{ \sqrt{g} } \sinh \left( \frac{\xi}{4} \right) \\
\label{eq:xtrans5}
dx &=&  \frac{1}{ 4 \sqrt{g} } \cosh \left( \frac{\xi}{4} \right) d \xi.
\eea
Using  (\ref{eq:KBess}), (\ref{eq:xtrans4}) and (\ref{eq:xtrans5}) we re-write ((\ref{eq:main_integral3}), using the shorthand sh (ch) for sinh (cosh): 
\be
\biggl< x^{2n} \biggr> = \mathcal{N} \int_0^{\infty} \mathrm{sh}^{2n} (\xi /4) \cdot \mathrm{ch}(\xi /4) \exp \left[ -\frac{1}{8g} \mathrm{ch}(\xi) \right] d\xi 
\label{eq:main_integral4}
\ee
Note the normalization factor $\mathcal{N}=\mathcal{N}(n,g)$:
\[
 \mathcal{N} (n,g)= \frac{1}{Z(g)} \cdot \frac{e^{1/8g}}{\sqrt{4\pi g^{2n+1}}} =\left[ g^n \cdot K_{1/4} \left(\frac{1}{8g}\right) \right]^{-1}
\]
We now make use of
\[
\sinh^{2n} (x) = \left[ \cosh^2 (x) - 1 \right]^n = \sum_{k=0}^n  \binom{n}{k}(-1)^{n-k} \mathrm{ch}^{2k}(x) ,
\]
to transform (\ref{eq:main_integral4}) to
\be
\label{eq:main_integral5}
\biggl< x^{2n} \biggr> = \mathcal{N}  \sum_{k=0}^{n}  \binom{n}{k}(-1)^{n-k} \int_0^{\infty} \mathrm{ch}^{2k+1}(\xi/4) \exp \left[ -\frac{1}{8g} \mathrm{ch}(\xi) \right] d\xi 
\ee
We first use the trigonometric identity for odd powers of cosine \cite{Wolf2} and then apply Osborn's rule \cite{Wolf3} to convert the identity to the hyperbolic analogue:
\[
\cosh^{2k+1}(x) = \frac{1}{4^{k}} \sum_{m=0}^{k} \binom{2k+1}{m} \mathrm{ch}( (2k + 1 - 2m )\cdot x)
\]
Then we finish up (\ref{eq:main_integral5})
\bea
\biggl< x^{2n} \biggr> &=& \mathcal{N} \sum_{k=0}^{n}  \binom{n}{k}\frac{(-1)^{n-k}}{4^{k}} \sum_{m=0}^{k}\binom{2k+1}{m} \times \nn \\
				& & \int_0^{\infty} \mathrm{ch}\left(\frac{(2(k-m)+1)\xi}{4}\right) \exp \left[ -\frac{1}{8g} \mathrm{ch}(\xi) \right] d\xi \nn \\
				&=& \mathcal{N} \sum_{k=0}^{n}  \binom{n}{k}\frac{(-1)^{n-k}}{4^{k}} \sum_{m=0}^{k} \binom{2k+1}{m} K_{\frac{2(k-m)+1}{4}} \left( \frac{1}{8g} \right) \label{eq:main_soln3}
\eea
In general we see that (\ref{eq:main_soln3}) is of the form
\be
\biggl< x^{2n} \biggr> = \left[ (4g)^n \cdot K_{1/4}  \left( \frac{1}{8g} \right) \right]^{-1} \sum_{m=0}^n c_m^{(n)} \cdot  K_{\frac{2m+1}{4}},
\ee
where with a little work we see that the coefficients $c_m^{(n)}$ can be expressed as: 
\be
c_m^{(n)} = \sum_{k=m}^n (-1)^{n-k} 4^{n-k} \binom{n}{k} \binom{2k+1}{k-m}
\ee
The coefficients can be simplified when starting from the top in descending order - for example the first few entries are
\bea
c_n^{(n)} &=& 1 \\
c_{n-1}^{(n)}  &=& 1 - 2n \\
c_{n-2}^{(n)}  &=& n (2n-3) \\
c_{n-3}^{(n)}  &=& \frac{n (1 - 2n) (2n - 5)}{3}
\eea
The even moments ($2n$) for $n=1,2,3$ are then given by:
\bea
\biggl< x^{2} \biggr> &=& \frac{1}{4g} \left[ \frac{K_{3/4} \left( \frac{1}{8g} \right)}{ K_{1/4}  \left( \frac{1}{8g} \right)} -1 \right] \\
\biggl< x^{4} \biggr> &=& \frac{1}{16g^2} \left[ \frac{K_{5/4} \left( \frac{1}{8g} \right)-3K_{3/4} \left( \frac{1}{8g} \right)}{ K_{1/4}  \left( \frac{1}{8g} \right)} +2 \right] \\
\biggl< x^{6} \biggr> &=& \frac{1}{64g^3} \left[ \frac{K_{7/4} \left( \frac{1}{8g} \right) -5 K_{5/4} \left( \frac{1}{8g} \right)+9K_{3/4} \left( \frac{1}{8g} \right)}{ K_{1/4}  \left( \frac{1}{8g} \right)} -5 \right] 
\eea

\section{Discussion}
As I was unable to find a derivation of the closed form expressions for $\phi^4$-field theory in zero dimensions, I set about deriving the expression on my own. Along the way the trick I used to derived the expression also enables one to write a closed form expression for the moments. I hope these results can lead to further insights on resummation methods used in the perturbative approach explored in \cite{JCC, HK, VM, Yuk} and references therein. It would be interesting to see if the hyperbolic transformation has a role to play in other field theories. I also discovered an erratum in one of the quoted formulas \cite{ERD} for which the correction is mentioned in the appendix.

\appendix 

\section{Calculations} \label{sec:appA}
First we note that the transformation (\ref{eq:xtrans1}) leads to the infinitesimal change
\be
\label{eq:xtrans2}
dx =  \frac{\sqrt{2}\gamma}{4\beta}\cosh \left( \frac{\xi}{4} \right) d \xi.
\ee
When switching to the variable $\xi$, the limits in the integral in (\ref{eq:main_integral2}) are unchanged. Next we consider the exponent in (\ref{eq:main_integral2}):
\bea
2 \gamma^2 x^2 + \beta^2 x^4 &=& 2 \gamma^2 x^2  \cdot \left[1 + \frac{\beta^2}{2\gamma^2} x^2 \right] \nn \\
					&=& \frac{4\gamma^4}{\beta^2} \sinh^2 \left( \frac{\xi}{4} \right)  \cdot \left[1 + \sinh^2 \left( \frac{\xi}{4} \right) \right] \nn \\
					&=&  \frac{4\gamma^4}{\beta^2} \sinh^2 \left( \frac{\xi}{4} \right) \cosh^2 \left( \frac{\xi}{4} \right) \nn \\
					&=&   \frac{\gamma^4}{\beta^2}  \sinh^2 \left( \frac{\xi}{2} \right) \nn \\
					&=&  \frac{\gamma^4}{\beta^2}  \cdot \left[ \cosh^2 \left( \frac{\xi}{2} \right) - 1 \right] \nn \\
					&=& \frac{\gamma^4}{\beta^2}  \cdot \left[ \frac{1}{2} ( \cosh \xi - 1) - 1 \right] \nn \\
					&=& \frac{\gamma^4}{2\beta^2}  \cosh \xi - \frac{\gamma^4}{2\beta^2} . \label{eq:xtrans3}
\eea

\subsection{Erratum in Erd\'{e}lyi et. al.}
The formula in \cite[Ch. 4.5 No. 34 p. 147]{ERD} states that
\be
\label{eq:main_soln4e}
 \int_{0}^{\infty}  (2t)^{-3/4} e^{-2a^{1/2}t^{1/2}} e^{-pt} dt = \left( \frac{a}{2p} \right)^{1/2} \exp \left[ \frac{a}{2p} \right] K_{1/4} \left( \frac{a}{2p} \right)
\ee
We also note the conditions for (\ref{eq:main_soln4e}) are stated as $| \arg a | < \pi$ and Re $p>0$ \cite{ERD}.
Using the substitution $t = x^4$ transforms (\ref{eq:main_soln4e}) to:
\be
\label{eq:main_soln4e2}
 2^{5/4} \int_{0}^{\infty}  e^{-2a^{1/2}x^2 -px^4} dx 
 \ee
We see that the integral in (\ref{eq:main_soln4e2}) is equivalent to setting $\gamma^2 = a^{1/2}$ and $\beta^2=p$ in (\ref{eq:main_integral2}) and subsequently use (\ref{eq:main_soln2}) to see that the correct version of (\ref{eq:main_soln4e}) is
\be
\int_{0}^{\infty}  (2t)^{-3/4} e^{-2a^{1/2}t^{1/2}} e^{-pt} dt = \left( \frac{a}{2p^2} \right)^{1/4} \exp \left[ \frac{a}{2p} \right] K_{1/4} \left( \frac{a}{2p} \right)
\ee

\end{document}